\documentclass[twocolumn, prd, nofootinbib,
showpacs, preprintnumbers, amsmath, amssymb]
              {revtex4}

\usepackage{graphicx}

\begin{document}

\title{Cosmological interplay between general relativity and particle physics }

\author{ Michael~Maziashvili}
\email{mishamazia@hotmail.com} \affiliation{ Andronikashvili
Institute of Physics, 6 Tamarashvili St., Tbilisi 0177, Georgia \\
Faculty of Physics and Mathematics, Chavchavadze State University,
32 Chavchavadze Ave., Tbilisi 0179, Georgia }

\begin{abstract}

We clearly formulate and study further a conjecture of effective
field theory interaction with gravity in the cosmological context.
The conjecture stems from the fact that the melding of quantum
theory and gravity typically indicates the presence of an inherent
UV cutoff. Taking note of the physical origin of this UV cutoff,
that the background metric fluctuations does not allow QFT to
operate with a better precision than the background space
resolution, we conjecture that the converse statement might also
be true. That is, an effective field theory could not perceive the
background space with a better precision than it is allowed by its
intrinsic UV scale. Some of the subtleties and cosmological
implications of this conjecture are explored.

\end{abstract}





\maketitle

\section*{Introduction}

A fairly generic fact in describing of nature is that the low
energy behavior of a system is largely independent on the details
of what is going on in higher energy scales. For instance, the
actual physical theory based on quantized fields is an effective
description applicable at some energy scale. In cosmology we have
lots of particle physics involved at various energy scales. In
this regard it is important to look carefully at the coupling of
gravity with the effective field theory in the cosmological
context. The effective field theory description usually includes
characteristic UV energy scale $\Lambda$. This fact immediately
underlines that there is a certain amount of frame dependence in
any effective field theory description. For our consideration it
is cosmological frame, that is, a preferred frame in which the
Cosmic Microwave Background Radiation is spatially isotropic. On
the other hand, in the cosmological context we have a natural IR
scale set by the cosmological horizon. That is, motivated by
emphasis on the operational meaning we restrict our consideration
to a causally connected region set by the cosmological horizon
that can be accessible to an observer. As the effective field
theory with UV scale $\Lambda$ describes the physical processes
with maximum spatial resolution $\Lambda^{-1}$, we conjecture that
such a theory describing particle physics content over the horizon
scale perceives background space with a precision $\delta l
\lesssim \Lambda^{-1}$. That is, if the background space is
Minkowskian for instance, the effective field theory with IR scale
$l$ will perceive it up to the fluctuations $\delta g_{\mu\nu}
\sim 1/l\Lambda $. Otherwise stated, in the Minkowskian background
space the effective quantum field theory can not differentiate
between $\eta_{\mu\nu}$ and $\eta_{\mu\nu} + \delta g_{\mu\nu}$,
where $\eta_{\mu\nu}$ is a Lorentz metric and $\delta g_{\mu\nu}
\lesssim 1/l\Lambda$. The background metric fluctuations allowed
by the effective quantum field theory certainly carries some
energy density contributing to the vacuum energy. In the
cosmological context this energy density can be estimated as
follows \cite{Sasakura}. Throughout this paper we use the system
of units in which $\hbar = c = k_B = 1$. The minimal cell by which
effective quantum field theory "measures" the horizon volume $l^3$
is set by $\Lambda^{-3}$. Assuming the finiteness of universe's
age, $t$, the time-energy uncertainty relation tells us that the
spatial cell $\Lambda^{-3}$ will carry energy of the order
$E_{\Lambda^{-3}} \sim t^{-1}$. Thus the energy density of the
background metric fluctuations takes the form \cite{mazia}
\begin{equation}\label{vacuumenergy} \rho_{vacuum} \sim {\Lambda^3(t) \over
t}~.\end{equation} To illuminate this conjecture further let us
notice that well-known quantum gravity arguments indicate the
existence of a natural UV cutoff in nature. Such UV cutoff even
when it is not set immediately by the Planck mass ($m_P \simeq
10^{19}$\,GeV) is very high at the particle physics scale. Let us
sketch one of the discussions of this kind \cite{CKN}. For an
effective quantum field theory in a box of size $l$ with UV cutoff
$\Lambda$ the entropy $S_{QFT}$ scales as,
\[ S_{QFT} \sim l^3\Lambda^3~.\] That is, the effective quantum
field theory counts the degrees of freedom simply as the number of
cells $\Lambda^{-3}$ in the box $l^3$. Imposing black hole entropy
bound $ S_{QFT} \lesssim  S_{BH} \simeq (l / l_P )^2 $ one arrives
at the relationship between UV and IR cutoffs \cite{CKN}
\begin{equation}\label{BHentropybound} \Lambda  ~\lesssim ~{1  \over l_P^{2/3}\,
l^{1/3} } \,~.
\end{equation} Albeit in Eq.(\ref{BHentropybound}) $\Lambda \ll m_P$ whenever $l^{1/3} \gg l_P^{1/3}$,
still it is very high from the standpoint of particle physics we
have in early cosmology. For instance estimating the horizon
distance for radiation dominated epoch
\[a(t) \propto t^{1/2}~,~~~~~ l = a(t)\int\limits_0\limits^{t} {d\xi \over a(\xi)} = 2t~, \] from Eq.(\ref{BHentropybound})
one finds that at EW phase transition when the universe was about
$\sim 10^{-12}$\,sec old $\Lambda \sim 10^{9}$\,GeV. Clearly the
EW theory that represents an effective field theory description of
this phase transition is characterized by the EW scale $\sim
10^2$\,GeV and does not care about gravity as the dimensionless
gravitational coupling $G_NE^2$ at this energy scale ($E \sim
10^2$\,GeV) is negligibly small. In general particle physics
describing the cosmological plasma does not care abut gravity as
long as its temperature is much less than Planck energy $T \ll
m_P$. Before passing to the subject of our discussion let us
recall some of the lore concerning QFT vacuum energy.

\section*{Vacuum energy in view of particle physics}

Particle physics contributes to the vacuum energy in two different
ways \cite{Weinberg}. First, we have the divergent {\tt
Nullpunktsenergie} characteristic of generic quantum field
theories.  Second, according to the SM, and even more so in its
unified extensions, what we commonly regard as empty space is full
of condensates. On the quite general grounds, as long as QFT
respects Lorentz invariance, one infers that the vacuum energy
mimics the cosmological constant \cite{Zeldovich}, to wit
\begin{equation}\label{Lorentzinv} \langle 0|T_{\mu\nu}|0\rangle =
\langle 0|T_{00}|0\rangle \,\eta_{\mu\nu}~.\end{equation}
\newline
{\tt \bf Nullpunktsenergie -} The vacuum energy density defined as
a {\tt Nullpunktsenergie} appears to be infinite. However, the
infinity arises from the contribution of modes with very small
wavelengths and for we do not know what actually might happen at
such scales it is reasonable to introduce a cutoff and hope that a
more complete theory will eventually provide a physical
justification for doing so. Before going on let us make a brief
comment on the regularization of {\tt Nullpunktsenergie}. In
presence of UV cutoff it is customary to set the energy density
coming from {\tt Nullpunktsenergie}
 as $\sim
\Lambda ^4$ (one can do so simply on the dimensional grounds).
However, one should care the Eq.(\ref{Lorentzinv}) to be satisfied
\cite{eqofstate}. Regularizations of the {\tt Nullpunktsenergie}
which respect the Lorentz symmetry of the underling theory
disfavor its quartic dependence on the UV scale, but rather it
appears to depend quadratically on the UV scale, $\sim
m^2\Lambda^2$, where $m$ is the mass scale of theory
\cite{eqofstate}. This point has attracted little attention
hitherto for many authors still follow the old customary.
Returning to the main stream of reasoning let us notice that in
QFT the energy-momentum operator $T_{\mu\nu}$ (and correspondingly
the source of gravity $\langle 0|T_{\mu\nu}|0\rangle $) is not
uniquely defined because of operator ordering. In the framework of
QFT we are usually subtracting this (divergent) {\tt
Nullpunktsenergie} which is equivalent to the normal operator
ordering in $T_{\mu\nu}$. Or equivalently in the path integral
approach one observes that the equations of motion for matter
fields are invariant under the shift of the matter Lagrangian by a
constant that results in a new energy momentum tensor
\[T_{\mu\nu}~~\rightarrow ~~T_{\mu\nu} +
\mbox{const.}\,\eta_{\mu\nu} ~.\] Usually in the framework of QFT
the vacuum energy $H|0\rangle = E_0|0\rangle$ is treated as an
unphysical quantity that may be set arbitrarily\footnote{For a
crystal the {\tt Nullpunktsenergie} represents the vibration
energy of crystal molecules at a zero temperature that manifests
itself even at a finite temperature and has therefore quite
definite physical meaning, see for instance very readable popular
book by Kaganov \cite{Kaganov}.}. Let us notice that in curved
space-time one can still renormalize the {\tt Nullpunktsenergie}
and set it completely arbitrary like in the case of Minkowski
background \cite{Ford}.
\newline
{\tt \bf Vacuum condensates -} The origin of masses is a
fundamental problem in particle physics. Conventionally, for
generating masses to the particles a self-interacting scalar field
is introduced in the SM that acquires a non-vanishing vacuum
expectation value and breaks the electroweak symmetry down to the
electromagnetic one. The classical, zero-temperature Higgs
potential is of the form

\begin{equation}  V_{\mbox{cl}}(\phi) =
\mbox{const.} - \mu^2 \phi^2 + {\tt g} \,\phi^4~, \end{equation}
with $\mu^2,\,{\tt g} > 0$ (usually ${\tt g}$ is small enough as
to ensure the perturbative regime of the theory). The minimum of
$V_{\mbox{cl}}$ breaks the $Z_2$ symmetry as $\phi$ acquires a
non-vanishing vacuum expectation value $\langle \phi \rangle =
\mu/\sqrt{2\,{\tt g}}$.

\begin{figure}[t]

\includegraphics[width= 0.48 \textwidth]{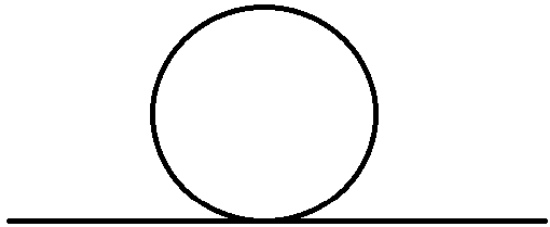}\\
\caption{Quadratically divergent loop giving rise to the thermal
mass. }
\end{figure}

At high temperature there is an additional contribution to the
scalar mass. Ignoring the gauge and fermion sectors, the effective
potential is approximately of the form \cite{Arnold}

\begin{equation} \label{oldeffectivepot} V_{\mbox{eff}}(\phi,~T) \simeq
\mbox{const.} + \left(- \mu^2  + {{\tt g}\, T^2 \over 2}
\right)\phi^2 + {\tt g}\, \phi^4~.
\end{equation} Diagrammatically, the thermal mass term in
Eq.(\ref{oldeffectivepot}) arises from the quadratically divergent
loop of Fig.1, where the UV divergence is cut off at momenta of
order $\sim T$. This quadratically divergent diagram evidently
provides a contribution of the order $\sim {\tt g}T^2$ for
temperatures large compared to the scalar mass \cite{Arnold}. The
addition of the thermal mass term above is responsible for
symmetry restoration at high temperature\footnote{The idea of high
temperature electroweak symmetry restoration was originally put
forward by Kirzhnits on the bases of similarity between Higgs
mechanism and the Ginzburg-Landau theory of superconductivity
\cite{Kirzhnits}.}. Namely for $T \gtrsim \mu/\sqrt{{\tt g}}$
effective potential (\ref{oldeffectivepot}) has a positive
coefficient for $\phi^2$ and then the minimum occurs at $\langle
\phi \rangle = 0$. If we adjust the $\mbox{ const.}$ in
Eq.(\ref{oldeffectivepot}) to get the zero effective potential for
$T \ll \mu/\sqrt{{\tt g}}$, then we have to put up with a very
large cosmological constant before the electroweak phase
transition, $V_{\mbox{eff}}(\phi = 0) \simeq \mu^4/{\tt g}$.
Nevertheless, as it was observed in \cite{BludRud}, this huge
cosmological constant is negligible compared to the radiation
energy present at that time, $T \sim \mu/\sqrt{{\tt g}}~
\Rightarrow  \rho_{rad} \sim \mu^4/{\tt g}^2$ . Thus, we see that
the vacuum energy density is suppressed by a large factor $1/ {\tt
g}$ compared to the radiation energy density and can be safely
ignored notwithstanding its large value.

The Higgs mechanism by means of which the standard models
particles (gauge bosons, leptons and quarks) acquire the mass has
little to do with the origin of mass of visible universe. The
visible world - stars, planets, galaxies, all bodies surrounding
us, are built from protons, neutrons and electrons. The
contribution of electrons to the mass of matter is negligibly
small (less  than $0.1$ percent). Therefore, in order to
understand the origin of mass of the observable world, it is
necessary to clarify the origin of nucleon mass. The nucleon
consists of $u$ and $d$ quarks. But the masses of $u$ and $d $
quarks are very small compared to the mass of nucleon, $m_u + m_d
\simeq 10$\,MeV and their contribution to the nucleon mass is
about $2$ percent. It can be shown that even in the formal limit
$m_u, m_d \to 0$, the nucleon mass remains practically unaltered.
The nucleon mass arises due to spontaneous violation of chiral
symmetry in QCD and can be expressed through the chiral symmetry
violating vacuum condensates \cite{Ioffe}.

In addition there are neutrino masses connected with the energy
scale beyond the SM. This energy scale ranges from $\sim$\,TeV to
the GUT scale $\sim 10^{16}$\,GeV with respect to the different
scenarios. The underlaying physics still awaits better
understanding, but in any case this energy scale does not appear
to be of great importance for what follows.

To summarize, the vacuum energy density coming from the SM
particle physics can be made consistent with the observed value of
dark energy density by adding a suitable counter term to $\langle
T_{\mu\nu}\rangle$

\[\langle T_{\mu\nu}\rangle~~\rightarrow ~~\langle T_{\mu\nu}\rangle +
\mbox{const.}\,\eta_{\mu\nu} ~,\] but certainly it does not
explain anything concerning the origin of dark energy. It merely
tells us that the SM particle physics does not contradict the
observed value of dark energy density.

\section*{Effective field theory }

The modern view is to regard our fundamental theories, the
standard model of particle physics and general relativity, as a
low energy effective theories. The renormalizability technically
corresponds to the possibility of sending the energy cutoff
$\Lambda$ of a system to infinity (while keeping all the physical
quantities finite). Physically this means that the theory can be
extrapolated to infinitely small distances without encountering
new microscopic structures. However, we have no good reason to
suspect that the effects of our present theory are the whole story
at the highest energies. Happily enough, we do not need to know
what is going on at all scales at once in order to figure out how
nature works at a particular scale. Effective field theory allows
us to make predictions at low energies without making unwarranted
assumptions about what is going on at high energies. Various
important energy scales of particle physics come into play in
cosmology \cite{partcosm}. Let us briefly sketch the interplay
between particle physics and cosmology in the early universe. The
melding of particle physics and cosmology leads to the expectation
that cosmic plasma in the early universe underwent a few phase
transitions related to the grand unified symmetry
breaking\footnote{Most likely the universe did not undergo the GUT
phase transition. The point is that the inflation energy scale,
$E_{inflation}$, is bounded from above by (non) observation of
tensor fluctuations of the cosmic microwave background radiation
(relict gravitational wave background) \cite{RubSazhVery}, with
the current limit being $E_{inflation} \lesssim 10^{16}$\,GeV
\cite{Spergel}.}, to the electroweak symmetry breaking, during
which SM particles acquire the mass, and chiral symmetry breaking
of strong interaction. Then the evolution of the early universe
proceeds according to known high energy physics. During the early
evolution of the universe the temperature and density of the
cosmic plasma are very high and respectively collisions are
exceedingly frequent to keep different species of particles in
thermal equilibrium with each other. As the universe expands and
cools, the reaction rates begin to lag behind the expansion rate
and various particle species drop out of equilibrium, or as it is
said freeze out/decouple. Being direct experimental signature, it
is important to study the relic abundance of decoupled species.
When the temperature of the universe drops down to $T\sim 1$\,MeV
the weak interactions become frozen, neutrinos decouple from the
rest of matter and shortly thereafter neutrons and protons cease
to inter-convert. Soon thereafter, as the temperature drops
somewhat below the nuclear binding energy $\sim 1$\,MeV, namely
around $T\sim 100$\,keV, Big-Bang nucleosynthesis begins. The
neutrons combine with protons into light nuclei, mostly helium-4,
but also deuterium, helium-3, lithium-7 and others. These elements
remain in the universe, so that their primordial abundance is
measurable today. Big-Bang nucleosynthesis ended when the universe
was about $t \sim 200$\,sec old. Much later occurred hydrogen
recombination at $T \sim 0.2$\,eV, $t\sim 3\cdot 10^5$\,yrs, after
that the cosmic plasma became electrically neutral and CMBR
decoupled from the rest of matter. Now turning to our discussion,
the question is to get good understanding of $\Lambda(t)$ that
enters the Eq.(\ref{vacuumenergy}). We said nothing about the dark
matter. If all the matter were baryonic, by noticing that gravity
is the dominant force for large scale structure of the universe,
one could find it well motivated to take $\Lambda(t) =
\Lambda_{QCD}$ after the QCD phase transition as this energy scale
is responsible for generating (most of) the mass of the baryonic
matter. That is, in this case one could find it convincing that
the effective field theory appropriate for large scale structure
of the universe should be that one describing the origin of the
nucleon mass. As it was mentioned in the previous section, the
mass of nucleon arises due to chiral symmetry breaking of strong
interaction the characteristic energy scale for which is set by
the $\Lambda_{QCD}$ \cite{Ioffe}. Indeed, by substituting
$\Lambda_{QCD} \simeq 170$\,MeV and $t_0\simeq 10^{60}t_P$ in
Eq.(\ref{vacuumenergy}), one gets pretty good value for observed
dark energy density. Such dark energy decays linearly with time
and thereby can not spoil the successes of early cosmology.
Assuming that this energy component dominates presently  \[ H^2 =
{8\pi \over 3m_P^2} \rho_{vacuum}~,  \] the equation of state can
be simply estimated by using energy-momentum conservation
\[ p = -{\dot{\rho} \over 3 H} - \rho~,
\] giving \begin{equation}\label{pressure}  p_{vacuum} \simeq \sqrt{ m_P^2\Lambda^3_{QCD}\over 24\pi t^3 } -
   {\Lambda^3_{QCD} \over t}~. \end{equation} The second term in Eq.(\ref{pressure}) becomes dominant for
   \[ t \gtrsim {m_P^2
   \over 24\pi \Lambda^3_{QCD} } \simeq
   10^{58}t_P~. \] So, this dark energy exhibits a negative
   pressure just recently.  This scenario was proposed in
   \cite{mazia}. Let us notice that linearly decaying dark energy has been proposed earlier in \cite{Zee}
   on the bases of Dirac's large number hypothesis.

   Now, let us summarize the main lines of our discussion and take
   note of the fact that the matter content of the universe is
   dominated by the dark matter.

\section*{Conclusion }

As is well known the melding of quantum theory and gravity is
marked with an intrinsic UV cutoff. Namely, the combination of
relativistic and quantum effects implies that the conventional
notion of distance breaks down the latest at the Planck scale.
Physically the emergence of this UV cutoff is understood as a
result of background metric fluctuations. The coupling of gravity
with QFT implements this UV cutoff into QFT. While the concrete
realization of this picture may involve many specifics, see for
instance well known example of this kind implemented through the
generalized uncertainty relations \cite{KM}, physically it is this
fluctuation term that renders the UV cutoff of QFT bounded from
above. That is, UV scale of QFT can not be greater that the
resolution of the background space. The question that immediately
occurs is, is not the converse statement true? That is, can the
effective field theory describing the processes with spatial
resolution $\Lambda^{-1}$ see the background space to a better
accuracy than $\Lambda^{-1}$? This is the conjecture put forward
in \cite{mazia}. This question is not only interesting in its own
right; it could also cast new light on dark energy. Two important
questions that immediately occur are as follows. First, effective
field theory is characterized with some energy scale, thereby
reflecting appreciable amount of frame dependance. In our
discussion we assume the cosmological frame defined by the Cosmic
Microwave Background Radiation. Second question is to get good
understanding of $\Lambda(t)$ in the cosmological context. The
dominant force for large scale structure of the universe is
gravity. Presently the matter outweighs the radiation by a wide
margin. The most natural assumption would be that the effective
field theory appropriate for large scale structure of the universe
should be that one describing the origin of the mass of matter.
The mass of the visible matter we see around us and are part of
comes overwhelmingly from nucleons. Operating simply by the
visible matter one would find it motivated to hold $\Lambda(t)
\simeq \Lambda_{QCD}$ after the QCD phase transition. But visible
matter contributes to the energy budget of the universe about 5
percent while dark matter is about 25 percent. Albeit they are of
the same order, matter in the universe is more dark than visible.
Therefore, from our discussion one finds it reasonable to insert
in Eq.(\ref{vacuumenergy}) the energy scale appropriate to the
dark matter (rather then visible matter)
\[\rho_{vacuum} \simeq {\Lambda^3_{DM}\over t}\] The question of energy scale $\Lambda_{DM}$
appropriate to an effective field theory describing the origin of
mass(es) of dark matter particle(s) is not yet understood well.
However, if the presented conjecture really works, the black hole
energy bound on Eq.(\ref{vacuumenergy}) for present value of
horizon distance $l_0\sim 10^{60}l_P$, that is to require $l_0
\gtrsim l_P^2l_0^3 \rho_{vacuum} $, puts an upper limit on
$\Lambda$ to be of the order of $\sim 1$\,GeV. That is, in this
case $\Lambda_{DM}$ could not be far above the $\Lambda_{QCD}$.
For the moment we can say nothing definitely about $\Lambda_{DM}$,
but certainly it would be nice this energy scale to be not very
far from $\Lambda_{QCD}$.

\vspace{0.5cm}
 This work was supported by the \emph{CRDF/GRDF} and the
\emph{Georgian President Fellowship for Young Scientists}.

\end{document}